\documentstyle[aps,twocolumn]{revtex}
\input epsf
\tightenlines
\begin{document}
\draft
\title{
{\bf Evolution of strangeness in  
equilibrating and expanding quark-gluon plasma}}
\author
{Dipali Pal\thanks{Present Address: Particle Physics Department,
Weizmann Institute of Science, Post Box 26, Rehovot 76100, Israel},
Abhijit Sen, Munshi Golam Mustafa}
\vskip 0.4in
\address
{Theory Group, Saha Institute of Nuclear Physics, 1/AF Bidhan Nagar, 
Kolkata 700 064, India.}
\vskip 0.4in
\author
{Dinesh Kumar Srivastava}
\vskip 0.4in
\address
{ Variable Energy Cyclotron Centre, 1/AF Bidhan Nagar, Kolkata
700 064, India.}
\date{\today}
\maketitle

\begin{abstract}


We evaluate the strangeness production from equilibrating 
and transversely expanding quark gluon plasma which may be created in the wake 
of relativistic heavy ion collisions. We consider  boost invariant 
longitudinal and cylindrically symmetric transverse expansion of a gluon 
dominated partonic plasma, which is in local thermal 
equilibrium. Initial conditions obtained from the self screened parton 
cascade model are used. 
We empirically find that the final extent of the partonic
equilibration rises almost linearly with the square of the initial
energy density. This along with the corresponding variation with the
number of participants may help us distinguish between various models of 
parton production.

\end{abstract}


\vskip 0.25in

PACS: 12.38.Mh, 24.85.+p, 25.75.-q, 13.85.Qk

\narrowtext
\vskip 0.25in

\section{\bf INTRODUCTION}

The study of ultra-relativistic heavy ion collision has entered a new
era with the successful commissioning of the Relativistic Heavy Ion
Collider at Brookhaven.  This provides an
opportunity to verify the possible occurrence of a phase transition from
hadronic matter to deconfined quark matter, where partons are the basic 
degrees of freedom. At the (lower) SPS energies, an
enhanced production of strangeness, considered to be one of the more
robust signatures of quark-hadron phase transition has already been
observed~\cite{sig,sig1,sig2,sig3}. The initial temperatures likely to be 
attained at RHIC and
the LHC  are expected to be much larger.  A natural question would now
be: how quickly is the strangeness equilibrated, if at all? 

In heavy-ion collisions strangeness is produced abundantly through the partonic
interactions if the temperature $T\ge 200$ MeV, the mass threshold of this 
semi-heavy flavour.  The extent of its  equilibration would however
depend upon such details as the  thermal and chemical 
evolution of the partonic system and the life-time of the
hot deconfined phase.
 It has
recently been shown that the chemical equilibration of the light flavours
and the gluons slows down due to the radial expansion\cite{chem2,duncan}.
 It should then be expected that the
extent of strangeness-equilibration can also be affected if
allowances are made for the likely radial expansion of the plasma. The
present work attempts to get answers to this and related questions.
We limit our discussions to the production and evolution of strangeness
during the deconfined phase, whose initial conditions are taken from the
Self Screened Parton Cascade (SSPC) model~\cite{sspc}, which has
formed the basis of a large number of related studies in recent times.
An early work in this direction used the initial conditions obtained
from the HIJING model and considered only a longitudinal 
expansion~\cite{levai}. In the following we closely follow this treatment
and extend it to include transverse expansion as well.

Our paper has been organized as follows: Section II describes briefly
the basic equations of the hydrodynamic and chemical evolution of the 
partonic gas through the partonic reactions in a (1+1) dimensional 
longitudinal expansion and a (3+1)
dimensional transverse expansion. 
A brief summary is given in section III.

\section{\bf HYDRODYNAMIC EXPANSION AND CHEMICAL EVOLUTION }

\subsection{Master Equations}
 
We start with the assumption that the early (semi)hard collisions among
partons produce a thermalized partonic plasma.
The high $p_T$ partons produced early in the collision,
then provide a colour screened environment for the production of
partons having low $p_T$ and the high density of the partons
launches the Landau Pomeranchuk Migdal (LPM) suppression mechanism to
eliminate the collinear singularity in parton fragmentation. 
This leads to the the so-called self screened parton cascade
model~\cite{sspc} and can be used to provide plausible initial
conditions for the system.  The  subsequent chemical equilibration
is then attained through reactions of the type 
$gg \leftrightarrow q\overline{q}$ and $gg \leftrightarrow ggg$.

 The evolution of the system is now controlled by the equation for
conservation of energy and momentum of an ideal fluid:
\begin{equation}
\partial_\mu T^{\mu \nu}=0 \; , \qquad
 T^{\mu \nu}=(\varepsilon+P) u^\mu u^\nu + P g^{\mu \nu} \, ,
\label{hydro}
\end{equation}
where $\varepsilon$ is the energy density and $P$ is the pressure measured
in the rest frame of the fluid. The four-velocity vector $u^\mu$ of 
the fluid satisfies the constraint $u^2=-1$.

We assume that the distribution functions for partons can be scaled
through equilibrium distributions as
\begin{equation}
f_j(E_j,\lambda_j)= \lambda_{j} {\tilde f}_j(E_j) \ \ , \label{befd}
\end{equation}
where ${\tilde f}_j(E_j)=({e^{\beta{E_j}}\mp 1})^{-1}$
is the BE (FD) distribution for gluons
(quarks), and $\lambda_j$ ($j=g, \ u, \ d, \ s$)  are the nonequilibrium
fugacities, $E_j=\sqrt{p_j^2+m_j^2}$, and $m_j$ is the mass of the parton.

Now one can write the number density, energy density and pressure for 
a partially equilibrated  multi-component partonic plasma~\cite{levai}
\begin{eqnarray}
n &=& n_g  +  \sum_i \left (n_{i}+n_{\bar i}
\right ) = \left ( \lambda_g a_1 + \sum_i \lambda_i b_1(x_i) \right ) T^3,
 \nonumber \\ 
\varepsilon &=& \varepsilon_g  +  \sum_i \left (\varepsilon_{i}+
\varepsilon_{\bar i}
\right ) = \left ( \lambda_g a_2 + \sum_i \lambda_i b_2(x_i)\right ) T^4, 
\nonumber \\ 
P &=& P_g  +  \sum_i \left (P_{i}+P_{\bar i} \right ) \nonumber \\
& =&  \frac{1}{3}\left ( \lambda_g a_3 + \sum_i \lambda_i b_2(x_i)
b_3(x_i) \right ) T^4,  \label{eos}
\end{eqnarray}
where $a_1=16\zeta(3)/\pi^2$ and $a_2=a_3=8\pi^2/15$ and
\begin{eqnarray}
b_1(x_i)&=&2 {d_i \over {2 \pi^2}} \cdot x_i^3
\sum_{n=1}^\infty (-1)^{n+1} {1 \over {n x_i}} K_2(n x_i), \nonumber \\
b_2(x_i)&=&2 {d_i \over {2 \pi^2}} \cdot x_i^4
\sum_{n=1}^\infty (-1)^{n+1} \left[
{3 \over {(n x_i)^2}} K_2(n x_i) \right. \nonumber \\
&& \ \ \ \left. +
{1 \over {(n x_i)  }} K_1(n x_i) \right], \nonumber \\
b_3(x_i) &=&
{ {  \sum_{n=1}^\infty (-1)^{n+1} {1 \over {(n x_i)^2}} K_2(n x_i) }
\over  { \sum_{n=1}^\infty (-1)^{n+1} \left[
{1 \over {(n x_i)^2}} K_2(n x_i) +
{1 \over 3} {1 \over {(n x_i)  }} K_1(n x_i)  \right]  }}, \nonumber \\
\label{a3}
\end{eqnarray}
for $i= u, \ d, \ s$.  We also have 
the colour and spin degeneracy $d_i=3\times 2$ and  $x_i=m_i/T$, where 
$m_i$ is the 
mass of the quark and $K$'s are modified Bessel functions. We take strange
quark mass as 150 MeV. For 
massless quarks these expressions simplify considerably and we have
 $b_1(0)= 2\cdot 9\zeta(3) /2\pi^2$, 
$b_2(0)=2\cdot 7\pi^2/40$ and 
$b_3(0)=1$. We further assume that 
$\lambda_i=\lambda_{\bar i}$, which should be valid for
negligible  net-baryonic density. This  should be a reasonable
approximation at RHIC and LHC  energies.
The speed of sound ($c_s$) can be obtained from
\begin{equation}
c_s^2 = \frac{dP}{d\varepsilon}.
\end{equation}
We found  it to be close to $1/\sqrt{3}$, as we confine ourselves to $T\ge$ 
200 MeV. We must add that several lattice QCD evaluations 
suggest~\cite{laeh} that
$\Delta=\varepsilon -3p \ge 0$ so that $c_s^2 < 1/3$. For such a situation
the cooling of the plasma would be slower~\cite{dipali} and thus a much
larger time would be available for equilibration.

We solve the hydrodynamic equations (\ref{hydro}) with the assumption
that the system undergoes a boost invariant longitudinal expansion along
the $z$-axis and a cylindrically symmetric transverse expansion \cite{vesa}.
It is then sufficient to solve the problem for $z=0$.

The chemical equilibration of the species $j$ is governed by the master equation\begin{equation}
\partial_\mu\left (n_ju^\mu\right ) = R_j(x) \ \ , \label{maschem}
\end{equation}
where $R_j$ are the rates  which propel the system towards chemical
equilibration. The system would be in chemical equilibrium when
$\lambda_j \equiv 1$ so that  $R_j(x)=0$.
As mentioned earlier, the dominant chemical reactions through which the
chemical equilibration
proceed~\cite{chem1} are $gg \leftrightarrow ggg$ and
$gg \leftrightarrow i\bar{i}$. Radiative processes involving quarks have 
substantially smaller cross sections in perturbative QCD, and quarks are less
abundant than gluons in the initial phase of the chemical evolution of the
parton gas. Other elastic scattering processes ensure maintenance of thermal
equilibrium.

Under these assumptions the
  master equations for different species become~\cite{chem1}
\begin{eqnarray}
\partial_\mu (n_g u^\mu)&=&(R_{2 \rightarrow 3} -R_{3 \rightarrow 2})
                    - \sum_i \left ( R_{g \rightarrow i}
                       - R_{i \rightarrow g} \right ) \, , \nonumber\\
\partial_\mu (n_i u^\mu)&=&\partial_\mu (n_{\bar{i}} u^\mu)
                     =  R_{g \rightarrow i}
                       - R_{i \rightarrow g},
\label{master1}
\end{eqnarray}
in an obvious notation.

The gain and loss term for the gluon fusion
process, $gg\leftrightarrow i{\bar i}$ can be written as
\begin{eqnarray}
R_{g \rightarrow i}-R_{i \rightarrow g} &=&\left( \lambda_g^2 -\lambda_i
\lambda_{\bar i} \right ) \frac{1}{2} \int \frac{d^3p_1}{(2\pi)^3 2E_1}
\int \frac{d^3p_2}{(2\pi)^3 2E_2} \nonumber \\
&&\int \frac{d^3p_3}{(2\pi)^3 2E_3} \int \frac{d^3p_4}{(2\pi)^3 2E_4}
{\tilde f}_g(p_1){\tilde f}_g(p_2) \nonumber \\
&&(2\pi)^4\delta^4\left (p_1+p_2-p_3-p_4\right ){\mathbf \sum} |{\cal M}_{gg
\rightarrow i{\bar i}}|^2, \nonumber \\ 
&& \label{fusrate}
\end{eqnarray}
where we have used the unitary relation $|{\cal M}_{gg \rightarrow 
i{\bar i}}|^2 = |{\cal M}_{i{\bar i} \rightarrow 
gg}|^2$. The above integral can be written as free space cross section for 
fusion process folded with the distributions for initial particles as
\begin{eqnarray}
I_{gg \rightarrow i{\bar i}} = \frac{1}{2} \int \frac{d^3p_1}{(2\pi)^3 }
\int \frac{d^3p_2}{(2\pi)^3 } 
\left [ \sigma_{gg\rightarrow i{\bar i}}v_{12}\right ]
{\tilde f}_g(p_1){\tilde f}_g(p_2)  \ , \label{fusrate1}
\end{eqnarray}
with the cross section given by
\begin{eqnarray}
d\sigma_{gg\rightarrow i{\bar i}}&=& \frac{1}{v_{12} \ 2E_1 \ 2E_2}
\int \frac{d^3p_3}{(2\pi)^3 2E_3} \int \frac{d^3p_4}{(2\pi)^3 2E_4}
\nonumber \\
&&(2\pi)^4\delta^4\left (p_1+p_2-p_3-p_4\right ){\mathbf \sum} |{\cal M}_{gg
\rightarrow i{\bar i}}|^2 \ , 
\label{fuscsec}
\end{eqnarray}
where $v_{12} = |{\mathbf v_1} - {\mathbf v_2}|$, the relative velocity 
between the initial particles.
Following Ref.~\cite{chem1} the Eq.(\ref{fusrate1})
can be written
\begin{equation}
I_{gg \rightarrow i{\bar i}}=\frac{1}{2} \sigma_2^i {\tilde n}_g^2
\end{equation}
 so that the rates given in Eq.(\ref{fusrate}) become 
\begin{eqnarray}
R_{g \rightarrow i}-R_{i \rightarrow g}&=&\frac{1}{2}\sigma_2^i n_g^2 
\left(1-\frac{\lambda_i^2 } {\lambda_g^2}\right) \nonumber \\
&=&R_2^i n_g
\left(1-\frac{\lambda_i^2} {\lambda_g^2}\right). \label{fg}
\end{eqnarray}
The net rate for the process $gg\leftrightarrow ggg$ 
can be written as
\begin{eqnarray}
R_{2 \rightarrow 3} -R_{3 \rightarrow 2}&=& (\lambda_g^2-\lambda_g^3)
\frac{1}{2} \int \frac{d^3p_1}{(2\pi)^3 \ 2E_1}
\int \frac{d^3p_2}{(2\pi)^3 \ 2E_2} \nonumber \\
&&\int \frac{d^3p_3}{(2\pi)^3 \ 2E_3}\int \frac{d^3p_4}{(2\pi)^3 \ 2E_4}
\int \frac{d^3p_5}{(2\pi)^3 \ 2E_5} \nonumber \\
&& \sum \left |{\cal M}_{gg\to ggg}\right |^2 f_g(p_1) f_g(p_2) 
(2\pi)^4 \nonumber \\
&&\delta^4(p_1+p_2-p_3-p_4-p_5). \label{gm0}
\end{eqnarray}
Similarly the integral in Eq.(\ref{gm0}) can also be written in 
factorised form~\cite{chem1}  
\begin{eqnarray}
R_{2 \rightarrow 3} -R_{3 \rightarrow 2}&=& \frac{1}{2}\sigma_3n_g^2
(1-\lambda_g)\ \nonumber \\
&=& R_3 n_g \left(1-\lambda_g\right)
. \label{gm1}
\end{eqnarray}
The density weighted rates in Eqs.(\ref{fg},\ref{gm1}) are defined as 
\begin{equation}
R_3 = \textstyle{{1\over 2}} \sigma_3 n_g, \quad
R_2^i = \textstyle{{1\over 2}} \sigma_2^i n_g ,  \label{eq16}
\end{equation}
where the thermally averaged and velocity weighted cross section are
\begin{equation}
\sigma_3 = \langle\sigma_{gg\to ggg}v_{12}\rangle, \quad \sigma_2^i =
\langle \sigma_{gg\to i{\overline{i}} }v_{12}\rangle , \label{eq10}
\end{equation}
where superscript $i$ in $R_2$ and $\sigma_2$ denotes to the fusion process
for a given flavour. We give the details of these calculations in the 
next sub-section. In calculating the rates, $R_2^i$ and $R_3$, we 
have also considered
the temperature dependence of the strong coupling constant as
\begin{equation}
 \alpha_s(T) = \frac{12\pi}{(33-2\times 3)\ln\left (Q^2/\Lambda_0^2\right)} ,
\label{coupling}
\end{equation}
with the cut-off $\Lambda_0=300 $ MeV.
The scale, $Q=2\pi T$, is derived by comparing the inverse of bare gluon 
propagator in the imaginary time formalism with the static longitudinal 
propagator~\cite{rafe2}.

Eq.(\ref{master1}) can now be simplified
\begin{eqnarray}
\frac{1}{n_g}\partial_t (n_g \gamma)&+&\frac{1}{n_g} 
\partial_r\left (n_g \gamma v_r\right )+\gamma
\left (\frac{v_r}{r} + \frac{1}{t}\right ) 
\nonumber \\
  &=& R_3 \left (1- \lambda_g\right ) 
        - 2\sum_iR_2^i \left ( 1-\frac{\lambda_i^2}
        {\lambda_g^2 }\right ) \, , \nonumber\\
\frac{1}{n_i}\partial_t (n_{i} \gamma)
&+& \frac{1}{n_i}\partial_r\left (n_{i} \gamma v_r\right )\ \ +  
\gamma \left (\frac{v_r}{r} + \frac{1}{t}\right ) \ \ \ \ \ \ \ \ \ \ \
 \nonumber \\
&=& R^i_2\frac{n_g}{n_i} \left ( 1-\frac{\lambda_i^2}
        { \lambda_g^2}\right ) \, ,
\label{mastf}
\end{eqnarray}
where $v_r$ is the transverse velocity and $\gamma=1/\sqrt{1-v_r^2}$. 

If we assume the system to undergo a purely longitudinal boost-invariant 
expansion, Eq.~(\ref{hydro}) reduces to the well-known relation~\cite{bjor}
\begin{equation}
{d\varepsilon\over d\tau} + {\varepsilon+P\over\tau} = 0. \label{eq12}
\end{equation}
where $\tau$ is the proper time.
Using Eqs.~(\ref{eos}) in Eq.~(\ref{eq12}), one can obtain the 
ultrarelativistic equation of motion~\cite{levai} as 
\begin{eqnarray}
 \left[ {{\dot{\lambda}_g} \over \lambda_g} + 4 {{\dot T} \over T} +
{4\over 3} {1 \over \tau} \right] \varepsilon_g +
\sum_i \left[ {{\dot{\lambda}_{i}} \over \lambda_{i}} + 4   {{\dot T} \over T}
{\cal D}_\varepsilon(x_i) \right. \ \ \ \ \ \ \ \ && \nonumber \\
\left. + {1 \over \tau} \left( 1 + {1\over 3} b_3(x_i) \right) \right] 
\varepsilon_i = 0 \ \ &&  \label{eq13}
\end{eqnarray}
where
\begin{eqnarray}
{\cal D}_\varepsilon(x_i) &=&
 \left\{   \sum_{n=1}^\infty (-1)^{n+1} \left[
{1 \over {(n x_i)^2}} K_2(n x_i) +
{5 \over{12}} {1 \over {(n x_i)}} \right.\right. \nonumber \\
&\times&\left.\left. K_1(n x_i) +  {1 \over{12}} K_0(n x_i) \right] \right\} 
\times \left\{ \sum_{n=1}^\infty (-1)^{n+1} \right. \nonumber \\
&\times& \left. \left[ {1 \over {(n x_i)^2}} K_2(n x_i) 
+ {1 \over 3} {1 \over {(n x_i)}}   K_1(n x_i) \right]  \right\}^{-1}  
 \label{a5}
\end{eqnarray}
which reduces to 
${\cal D}_\varepsilon(0) =b_3(0)=1$ for massless quarks (antiquarks).

Now the master equations~(\ref{maschem}) can be written as

 \begin{eqnarray}
\frac{\dot\lambda_g}{\lambda_g} + 3\frac{\dot T}{T} + \frac{1}{\tau}=
R_3 (1-\lambda_g)&-&
2 \sum_i R_2^i \left(1- \frac{\lambda_{i}^2}{\lambda_g^2} \right) \ ,
\label{eq14}
\end{eqnarray}
and
 \begin{eqnarray}
\frac{\dot\lambda_{i}}{\lambda_{i}} + 3\frac{\dot T}{T} {\cal D}_n(x_i)
+ \frac{1}{\tau}& =&
R_2^i {n_g\over n_{i}} \left( 1 - \frac{\lambda_{i}^2}{\lambda_g^2} \right)
\ \ ,
\label{eq15}
\end{eqnarray}
where 
\begin{equation}
{\cal D}_ n(x_i) =
 { {  \sum_{n=1}^\infty (-1)^{n+1} \left[ {1 \over {(n x_i)}} K_2(n x_i)
      + {1 \over 3} K_1(n x_i) \right] }
  \over { \sum_{n=1}^\infty (-1)^{n+1} {1 \over {(n x_i)}}
       K_2(n x_i)  }} , \label{a6}
\end{equation}
and ${\cal D}_n(0)=1$, for massless quarks. We can easily see that 
Eqs.(\ref{mastf}) reduce to Eqs.(\ref{eq14},\ref{eq15}) if there is no
transverse expansion so that the radial velocity, $v_r=0$.
Now one can study evolution of the multi-component partonic plasma in terms
of the parton fugacities by solving the Eq.~(\ref{eq15}) 
for longitudinally and Eq.~(\ref{mastf}) for transversely expanding plasma 
with the parton chemical equilibration rates. 

\subsection{Partons Equilibration Rates:}
  
In this subsection  we briefly recall~\cite{levai,muller} the evaluation of
equilibration rates for the gluon fusion process ($gg\to{i\bar i}$) and gluon
multiplication process ($gg\to ggg$).
 
\subsubsection{Gluon Fusion}

The differential cross section given in Eq.(\ref{fuscsec}) can be 
written as
\begin{equation}
\frac{{d\sigma}_{gg\to i\bar i}}{dt} =\frac{1}{16\pi \xi(s,m_1^2,m_2^2)}
{\mathbf \sum}|{\cal M}_{gg\to{i\bar i}}|^2 \ , \label{dsdt}
\end{equation} 
where $\xi=  {s^2-2s(m_1^2+m_2^2)+(m_1^2 -m_2^2)^2}$, which reduces
to  $\xi =s^2$, for 
massless incident particles. The matrix element for the process,
$gg\to i\bar i$, can be obtained in terms of Mandelstam variable 
from Ref.~\cite{com} as
\begin{eqnarray}
{\mathbf \sum}|{\cal M}_{gg\to{i\bar i}}|^2&=&\pi^2\alpha_s^2 \left[ 
\frac{12}{s^2}(M_i^2-t)(M_i^2-u)\right. \nonumber \\
&+&\left.\frac{8}{3}\frac{(M_i^2-t)(M_i^2-u)-2M_i^2(M_i^2+t)}
{(M_i^2-t)^2}\right. \nonumber \\
&+&\left.\frac{8}{3}\frac{(M_i^2-t)(M_i^2-u)-2M_i^2(M_i^2+u)}
{(M_i^2-u)^2}\right. \nonumber \\
&-&\left.6\frac{(M_i^2-t)(M_i^2-u)+M_i^2(u-t)}
{s(M_i^2-t)}\right. \nonumber \\
&-&\left.6\frac{(M_i^2-t)(M_i^2-u)+M_i^2(t-u)}
{s(M_i^2-u)}\right. \nonumber \\
&-&\left.\frac{2M_i^2(s-4M_i^2)}{3(M_i^2-t)(M_i^2-u)}\right ] \ \ , 
\label{matrix}
\end{eqnarray}
in which $M_i$ is the current quark mass. However, for massless quarks 
($q=$ $u$ and $d$)
the above Eq.(\ref{matrix}) reduces to
\begin{eqnarray}
{\mathbf \sum}|{\cal M}_{gg \to {q\bar q}
}|^2&=&16 \pi^2\alpha_s^2 
\left[\frac{3}{4}\frac{ut}{s^2} + \frac{1}{6} \left ( \frac{u}{t}+\frac{t}{u}
\right )-\frac{3}{8} \right ] \ . \label{matrix0}
\end{eqnarray}
This together with Eq.(\ref{dsdt}) diverges logarithmically 
as $u, \ \ t\rightarrow 0$. We assume that this logarithmic divergence
for massless quarks can be regularized by assigning them mass given 
by the thermal mass~\cite{chem1}
 
\begin{eqnarray}
m_{u,d}^2 = \left [ \lambda_g + \frac{1}{2} \left ( \lambda_u
+\lambda_d \right ) \right ]\frac{4\pi}{9}\alpha_s T^2 \ , \label{thmass}
\end{eqnarray}
For $s$-quark no such approximation is necessary and we use Eq.(\ref{matrix})
with $m_s=150$ MeV.

Now integrating the  matrix 
element in Eq.(\ref{dsdt}) over the variable $t$, between the limits
\begin{equation}
t_{\pm} = M_i^2 -\frac{s}{2} \left [1 \pm \chi \right ]
\ \ , \label{limit}
\end{equation}
one obtains the total cross-section
\begin{eqnarray}
\sigma_{gg\to i\bar i}& = & \frac{\pi \alpha_s^2}{3s} \left [
\left (1 + \frac{4M_i^2}{s} + \frac{M_i^4}{s^2} \right )\log\frac{1+\chi}{1-
\chi} \right. \nonumber \\
&-&\left. \chi\left(\frac{7}{4}+ \frac{31}{4}\frac{M_i^2}{s}\right) \right ],
\end{eqnarray}
where $\chi = \sqrt {1-{4M_i^2}/{s}}$.

Now the thermally averaged, velocity weighted cross section is defined as
\begin{eqnarray}
\sigma_2^i &=& \langle \sigma_{gg\to i\bar i} v_{12} \rangle
= \frac{\int d^3p_1 d^3p_2 f_g(p_1)f_g(p_2)\sigma_{gg\to i\bar i} v_{12}} 
{\int d^3p_1 d^3p_2 f_g(p_1)f_g(p_2)}.
\end{eqnarray}
Using $v_{12}= \xi^{1/2}(s,0,0)/2p_1p_2$ and the thermal average of $s$
for a pair of gluons, $\langle s\rangle=18T^2$, we get
\begin{equation}
\sigma_2^i \approx \frac{9}{4}\left ( \frac{\zeta (2)}{\zeta (3)}\right )^2 
\left. \sigma_{gg\to i\bar i} \right |_{ \ M_i; \ \ \langle s\rangle =18 T^2}
. \label{sig2}
\end{equation}
Combining Eqs.(\ref{eq16}) and (\ref{sig2}) we finally obtain
the gluon fusion rate. 

\subsubsection{Gluon Multiplication}
Following  Ref.~\cite{chem1}
we also estimate the $\sigma_3$ for gluon multiplication process from the 
differential
cross-section given by~\cite{muller}
\begin{eqnarray}
\frac{d\sigma_3} {dq_\bot^2\ d^2k_\bot \ dy} &=& \frac{d\sigma_{\rm{el}}^{gg}}
{dq_\bot^2} \ \frac{dn_g} { d^2k_\bot \ dy}
\theta\left (\lambda_f- \frac{\cosh y}{k_\bot} \right ) \nonumber \\
&& \times \ \theta\left ({\sqrt s}-k_\bot {\cosh y}\right )\ \ , \label{sig3} 
\end{eqnarray}
where the first step function includes the approximate LPM 
suppression of the induced gluon and the second step
function accounts for energy conservation.  Here 
$k_\bot$ denotes the transverse momentum, $y$ is the rapidity of the
radiated gluon and $q_\bot$ corresponds the momentum transfer in the
elastic collisions. The infrared divergence associated with QCD radiation
is regularized by the LPM effect. However, Eq.~(\ref{sig3}) is still having
infrared singularities in both scattering cross sections and
radiation amplitudes associated with the gluon propagator.  
One can, approximately, control all these singularities 
using the Debye screening mass~\cite{muller}
\begin{equation}
\mu_D^2=\frac{6g^2}{\pi^2}\int_0^\infty kf_g(k)dk = 
4\pi\alpha_s T^2 \lambda_g \label{debye} \ \ .
\end{equation}  
Now the regularized gluon density distribution induced by a single
scattering is 
\begin{equation}
\frac{dn_g} { d^2k_\bot \ dy} = \frac{3\alpha_s}{\pi^2} 
\frac{q_\bot^2} {k_\bot^2 \left [ \left( {\mathbf k_\bot -\mathbf q_\bot} 
\right )^2 + \mu_D^2 \right ] }, 
\end{equation}
and the regularized small angle $gg$ scattering cross section is
\begin{equation}
\frac{d\sigma_{\rm{el}}^{gg}}{dq_\bot^2} = \frac{9}{4} \frac{2\pi\alpha_s^2}
{\left( q_\bot^2 +\mu_D^2 \right )^2} \ \ .
\end{equation}
The mean free path for the elastic scattering is obtained as~\cite{muller}
\begin{eqnarray}
\lambda_f^{-1} = \frac{9}{8} a_1 \alpha_s T \frac{1}
{1+8\pi \alpha_s\lambda_g/9}.
\end{eqnarray}
Integrating the $\phi$ part analytically we get the gluon multiplication
rate as
\begin{equation}
R_3 = \frac{32}{3a_1} \alpha_s T \lambda_g \left ( 1 + \frac{8}{9} 
a_1\alpha_s \lambda_g \right )^2 {\cal I}(\lambda_g) \ \ ,
\end{equation}
where 
\begin{eqnarray}
{\cal I}(\lambda_g) &=& \int_1^{{\sqrt s}\lambda_f} dx \int_0^{s/4\mu_D^2} dz
{\frac{z}{(1+z)^2}} \nonumber \\
&& \left\{ 
\frac {\cosh^{-1} \left ( \sqrt x \right )} 
{x \sqrt{\left [ x + (1+z)x_D\right ]^2 -4xzx_D}} \right. \nonumber \\
&& \left. + \frac{1}{s\lambda_f^2} 
\frac {\cosh^{-1} \left ( \sqrt x \right )}
{\sqrt{\left [ 1 + x(1+z)y_D\right ]^2 -4xzy_D}} \ 
\right \} ,
\end{eqnarray}
with $x_D= \mu_D^2\lambda_f^2$ and $y_D=\mu_D^2/s$.

In the following subsection we 
discuss the chemical evolution of different parton species.

\subsection{Evolution of the Multi-component Parton Plasma}

As mentioned earlier we shall be using the initial conditions obtained from
the self-screened parton cascade model~\cite{sspc}. Even though
they are well-known and used in a large number of studies we reproduce them 
in Table-I for the sake of completeness and an easy reference. 
We have chosen an initial fugacity for the strange quarks as half of 
that for the light quarks as in Ref.~\cite{levai}. This is consistent
with the assumption, which is often made, that the number of flavours is
$\approx 2.5$ if the mass of $s$ quark is taken as zero~\cite{chem1}. 
We show our results for the longitudinal expansion for RHIC (upper panel)
and LHC (lower panel) energies in Figure 1. 
As the additional parton production consumes energy, the temperature of 
the partonic plasma drops
considerably faster than the ideal Bjorken's scaling 
($T=T_0 (\tau_0/\tau)^{1/3}$, $T_0$ and $\tau_0$, respectively, are 
initial temperature and time of the parton gas) represented by the dashed
line. 
We see that like the light quarks and gluon, 
the production of strange 
quarks continues till late in the evolution. 
We further note that the extent of equilibration 
for the strange quarks in comparison to that for the light quarks
($\lambda_s/\lambda_{u,d}$) rises rapidly and once the temperature falls below
about $300$ MeV ($\sim \ 2m_s$ ) it gets more or less frozen by 
this time.
Thus we conclude that the equilibration of strangeness production
may imply the existence of quark matter at a temperature of more than 
about 300 MeV for a time of the order of a few fm/$c$. The other aspects
of variation of $\lambda_g, \ \ \lambda_{u,d}$ and $T$ have 
already been discussed by several authors~\cite{chem2,levai,chem1}.
We do note that the plasma is not fully equilibrated chemically at either
RHIC or LHC energies by the time the temperature drops below 200 MeV. We do
expect that additional quark production may occur due to gluon 
fragmentations during hadronisation~\cite{levai,stock,klaus} leading to a
chemically equilibrated hot hadronic matter.

Before giving our results for transverse expansion we need to specify
the profile of the initial energy density and the fugacity of the
partonic system. Following Ref.~\cite{duncan,chem3,bkp} we take the initial
energy density as
\begin{equation}
\varepsilon \left ( r, \tau_0 \right ) = \frac{3}{2} \varepsilon_0 
\left [ 1- \frac{r^2} {R^2} \right ]^{1/2} \Theta(R-r),
\end{equation}
where $R$ is the transverse dimension of the system,  $r$ is the radial
distance, and $\varepsilon_0$ is the  ``average'' initial
energy density (see Tab. I).
 The profile plays an important role in defining the boundary
of the hot and dense deconfined matter and affects the transverse expansion
through the introduction of pressure gradients.
We have further taken $\lambda_j(r,\tau_0)\propto \varepsilon ( r, \tau_0 )$
as before~\cite{duncan,chem3,bkp}. 
Any other variation will require an additional parameter.
We give our results for the radial variation of $\lambda_g$,
$\lambda_{u,d}$ 
and $\lambda_s/\lambda_{u,d}$ in Figs. 2 and 3 for RHIC and LHC energies, 
respectively, for various times along the
constant energy density contours with $\tau =N\tau_0$. Here $N$ is  
defined~\cite{chem2,chem3} through
$\varepsilon(r,\tau)=\varepsilon(r=0,\tau_0)/N^{4/3}$.

We see that the fugacities attain their highest values near $r=0$ 
and rise rapidly first and only slowly later in time. We also see
a result unique to chemical evolution with transverse expansion that
the fugacities may even start reducing towards the end of the QGP phase 
when the radial velocity (gradient) becomes very large 
(see Ref.~\cite{chem2} for explanation).

We can use our results for the radial variation of the energy density and 
the fugacities
obtained here to estimate the extent of partonic equilibration as a function
of the initial energy density. Comparing results for this from figures 4 and
5 we note that once the energy density is beyond about 20--40 GeV, the 
final fugacities for all the partons increase almost linearly with the
square of the energy density. This is a very interesting result.

Let us identify the ``local'' energy density at the radius $r$ with the 
average energy density attained in a non-central collision or with
a central collision involving lighter nuclei. Recall that the partonic
models suggest that the initial temperature attained in such collisions
through the production of minijets varies as $A^{1/6}$~\cite{hwa,wang}. 
This should then
imply that the extent of equilibration of strangeness
produced would rise as 
$N_{\mathrm {part}}^{4/3}$, if we can identify $A$ with  
$N_{\mathrm{part}}/2$, where $N_{\mathrm{part}}$ is the number of 
participants. In actual cases
the variation may be somewhat modified from this naive expectation due to
the considerations of shadowing and jet quenching~\cite{gyul}.

If on the other hand the energy densities etc. are decided by the
considerations of parton saturation~\cite{eskola} then
the initial temperature would vary as $\sim A^{0.126}$ and the energy 
density as $\sim A^{0.5}$. The results of Figs. 4 and 5 would then 
imply that the extent of strangeness equilibration should increase
 as the number of participants, $N_{\mathrm {part}}$ (see also
Ref.~\cite{khar} for a slight reformulation of the parton saturation
model).

 It is also
interesting to note that a recent analysis~\cite{jean}
 of centrality dependence of 
the extent of strangeness equilibration at CERN SPS energies in Pb+Pb
collisions gives a linear increase for this with the number of 
participants.                     

\subsection{Comparison with other Works:}

Rafelski and coworkers have studied the strangeness equilibration 
for more than two decades in various details. We shall refer to only a recent 
work by these authors~\cite{john}. In Ref.~\cite{john} the following
assumptions have been made which are at variance with our works:

\begin{enumerate}
\item The light quarks and gluons are assumed to have attained chemical
equilibrium ($\lambda_g = \lambda_u = \lambda_d=1$)
when the evolution of strangeness begins.
\item A Boltzmann approximation is used to describe the phase space 
distribution of $s$-quarks.
\item The flavour changing process, $q\bar q \leftrightarrow s\bar s$,
has been included. We have  neglected them assuming that the gluon fusion 
process $gg\leftrightarrow s\bar s$ dominates in a chemically 
undersaturated plasma.
\end{enumerate}

Now in the Boltzmann limit the evolution of the $s$-quark
fugacity can be written from (\ref{eq15}) as
\begin{eqnarray}
\frac{\dot{\lambda_s}}{\lambda_s}+3\frac{\dot T}{T}
\left [1+\frac{x_s}{3}\frac{K_1(x_s)}
{K_2(x_s)}\right ] + \frac{1}{\tau} =
R^s_2\frac{n_g}{n_s} \left (1-\frac{\lambda_s^2}{\lambda_g^2}\right ). 
\end{eqnarray}
This can be rewritten as
\begin{eqnarray}
{\dot T}{\tilde n}_s \left [ \frac{d\lambda_s}{dT} + \frac{\lambda_s}{T}
x_s\frac{K_1(x_s)}{K_2(x_s)}\right. &+& \left. 3 \frac{\lambda_s}{T} \right ] 
+ \frac{n_s}{\tau} \nonumber \\
&=& R^s_2n_g \left (1-\frac{\lambda_s^2}{\lambda_g^2}\right ) ,
\end{eqnarray}
which corresponds to Eq.(11) of Ref.~\cite{john}. We see that the
$3^{\rm{rd}}$ and $4^{\rm{th}}$ terms in the left hand side  here are 
absent in Eq.(11) of Ref.~\cite{john}, due to the above discussion.
One can thus see that the extent of strangeness equilibration attained 
in the work of Ref.~\cite{john} is much larger than in the 
present work. 

As a further check, we have repeated our calculations, assuming
$\lambda_g=\lambda_u=\lambda_d=1$ with the same energy density
as before, (see Table I) and further taking $\lambda_s(\tau_0)=0.2$
as in the work of Ref.~\cite{john}.
We found that now our parameter $\lambda_s$ rises to about 0.4 at
RHIC energies and up to 0.72 at LHC energies, for the case
of longitudinal expansion (see Fig.~6). When the transverse expansion
is allowed these numbers reduce to 0.66 at LHC energies and 0.38 at
RHIC energies, respectively.
 We thus realize that
the extent of equilibration of strangeness depends sensitively on the
initial conditions and also on the evolution mechanism.

Wong has studied~\cite{wong} the chemical equilibration of plasma
essentially in a formalism similar to that used in the present work, 
but only with longitudinal expansions. His values for $\alpha_s$ are
also much larger.

\section{SUMMARY}

We have studied the evolution and production of strangeness 
through the partonic interactions in a chemically equilibrating 
and expanding multi-component partonic gas with the initial conditions 
obtained from SSPC model.
We find that most of the strange quarks are produced when the temperature 
is still more than about 300 MeV ($\sim 2m_s$). Thus a chemically equilibrated 
plasma is expected to imply the existence of QGP phase for a
duration of several fm/$c$. We also find  approximately that the 
extent of strangeness equilibration 
rises linearly with the square of the initial energy density
within our approach. This may help us to obtain the scaling of the
initial energy density with the number of participants and distinguish
between the minijet and the partonic saturation models of parton 
production. It is rather interesting that the charged particle
multiplicity in Au + Au collision at $\sqrt{s_{NN}}=130$ GeV measured by 
the PHENIX collaboration~\cite{phen} shows a behaviour which is a superposition
of two terms, a linear increase with the number of participants
and a linear increase with the number of collisions (which varies
as $N_{\mathrm{part}}^{4/3}$).

\vskip 0.3in


\noindent {\bf Acknowledgment}

One of us (D.P.) acknowledges the hospitality of the
theory group of Saha Institute of Nuclear Physics during her short term 
visit when most of the work was done. We also thank X.-N. Wang for useful
correspondence.

\newpage

 \begin{table}
\caption{Initial conditions for the hydrodynamical expansion phase
in central collision of two gold nuclei at BNL RHIC and CERN LHC 
energies, respectively, from SSPC\protect\cite{sspc} given
in first and second row. Initial conditionis given in third and fourth
row are obtained with the same energy densities as before along with
$\lambda_u=\lambda_u=\lambda_g=1$.}
\begin{center}
\begin{tabular}{|l|c|c|c|c|c|c|}
\hline
& & & &&&\\
Energy &
$\tau_0$ & $T_0$ & $\lambda_g^{(0)}$&
$\lambda_{u(d)}^{(0)}$&$\lambda_s^{(0)}$ & $\epsilon_0$ \\
& & & &&&\\
 & (fm/$c$) & (GeV)&-&-&-& (GeV/fm$^3$) \\
& & & & &&\\ \hline
&&&&&&\\
 RHIC & 0.25 & 0.67 & 0.34 & 0.068&0.034 & 61.4  \\
& & & & &&\\
  LHC & 0.25 & 1.02 & 0.43 & 0.086&0.043 & 425 \\
& & & & &&\\
\hline
&&&&&&\\
 RHIC & 0.25 & 0.44 & 1.0 & 1.0&0.2 & 61.4  \\
& & & & &&\\
  LHC & 0.25 & 0.72 & 1.0 & 1.0&0.2 & 425 \\
& & & & &&\\
\hline
\end{tabular}
\end{center}
\end{table}

\begin{figure}
\epsfxsize=2.90in
\epsfbox{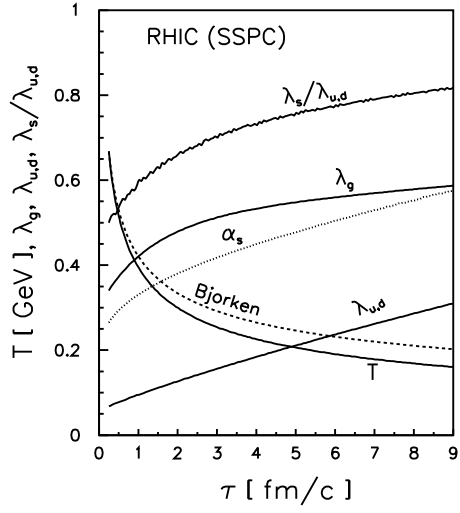}
\vskip -0.70in
\epsfxsize=2.90in
\epsfbox{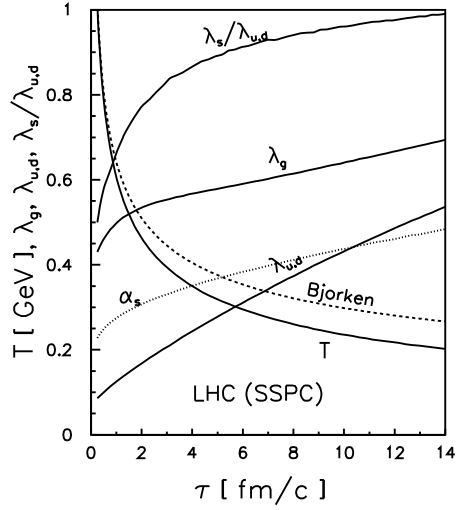}
\vskip -0.60in
\caption{Variation of temperature, coupling constant, gluon and quark 
fugacities with
proper time  for (1+1) dimensional hydrodynamic expansion with SSPC
initial condition for RHIC (upper panel) and LHC (lower panel) energies.}
\end{figure}

\newpage

\begin{figure}
\vskip -0.90in
\epsfxsize=2.90in
\epsfbox{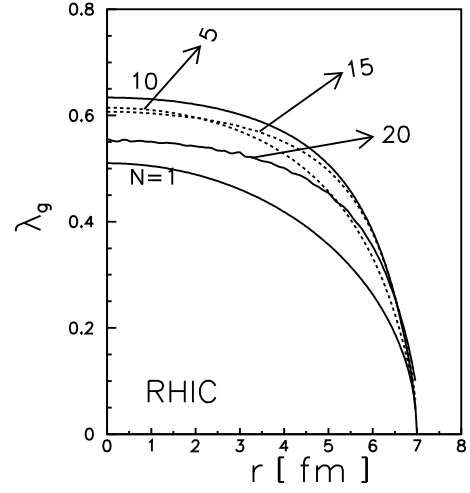}
\vskip -0.90in
\epsfxsize=2.90in
\epsfbox{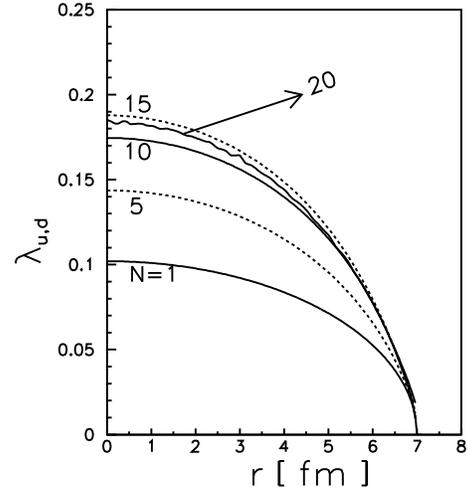}
\vskip -0.90in
\epsfxsize=2.90in
\epsfbox{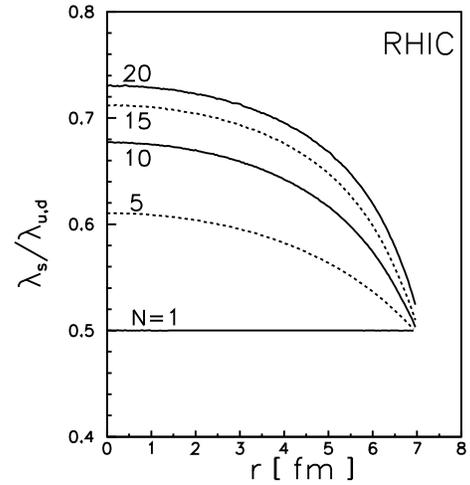}
\vskip -0.60in
\caption{ Variation of gluon (upper panel), massless quark (middle panel)
fugacities and ratio of strange to nonstrange quark fugacity (lower panel) 
with the transverse radius for RHIC energy at different times
\protect$\tau=N\tau_0$, along the constant energy density
contours defined by
 \protect$\varepsilon(r,\tau)=\varepsilon(r=0,\tau_0)/N^{4/3}$. }
\end{figure}

\begin{figure}
\epsfxsize=2.90in
\epsfbox{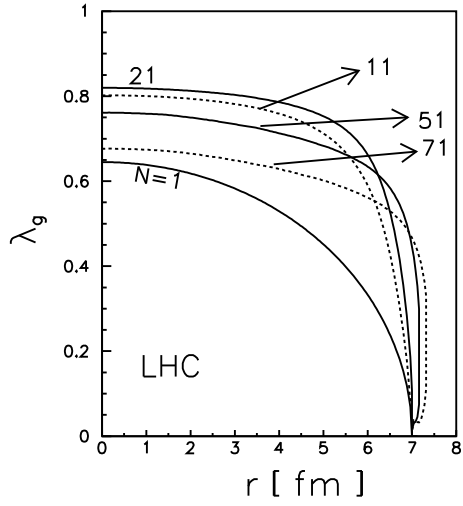}
\vskip -0.90in
\epsfxsize=2.90in
\epsfbox{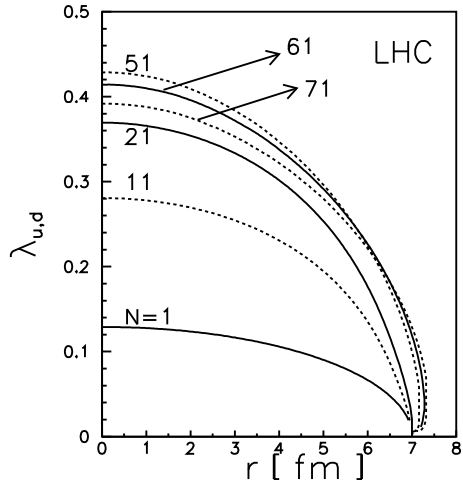}
\vskip -0.90in
\epsfxsize=2.90in
\epsfbox{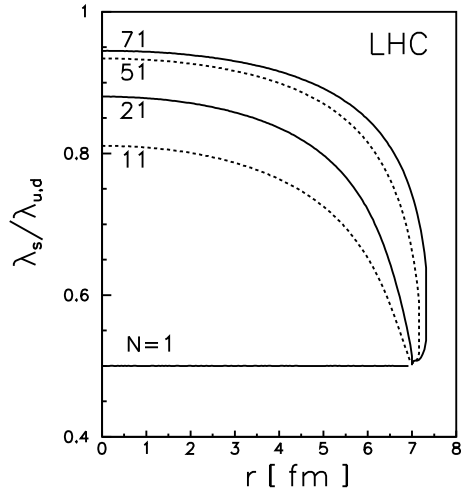}
\caption{Same as Fig. 2 for LHC energy.} 
\end{figure}

\begin{figure}
\vskip -0.90in
\epsfxsize=2.90in
\epsfbox{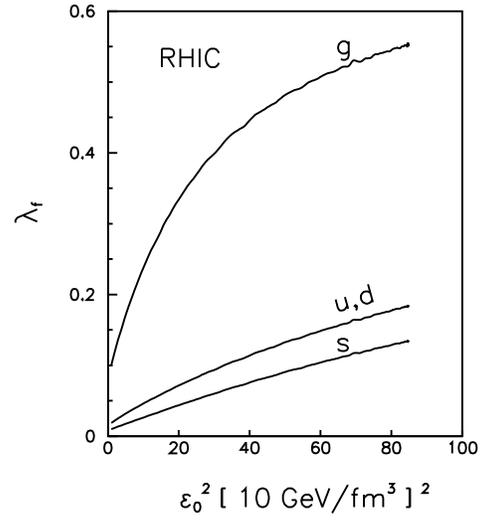}
\vskip -0.20in
\caption{Variation of the final fugacities with the initial energy density
for RHIC energy. Note the scale of the x-axis} 
\end{figure}

\begin{figure}
\epsfxsize=2.90in
\epsfbox{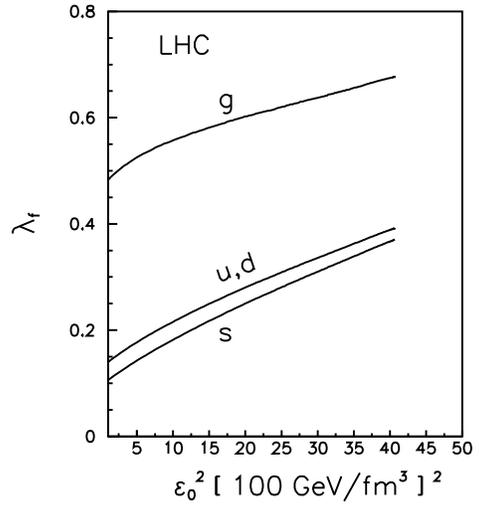}
\vskip -0.20in
\caption{Same as Fig. 4 for LHC energy. Note the scale of the 
x-axis.} 
\end{figure}


\begin{figure}
\epsfxsize=2.90in
\epsfbox{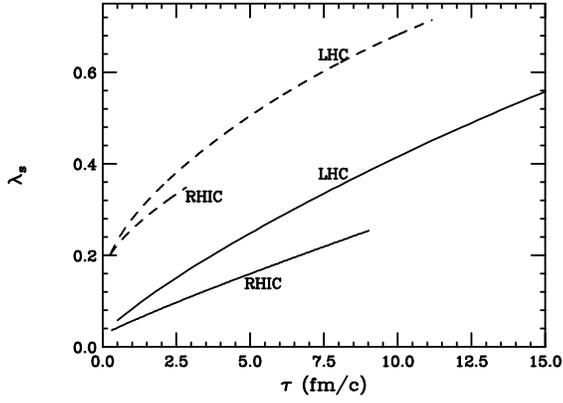}
\vskip +0.20in
\caption{Sensitivity of the strangeness equilibration
to the initial conditions for longitudinal expansion. The solid curves 
give the results corresponding
to Fig.~1 given earlier, while the dashed curves correspond to initial
condition which have the same energy densities as before, but have
$\lambda_u=\lambda_u=\lambda_g=1$ and $\lambda_s=0.2$, as given in
Table~I. 
} 
\end{figure}

\end{document}